\documentclass[twocolumn,showpacs,preprintnumbers,amsmath,amssymb]{revtex4}

\usepackage{graphicx}
\usepackage{dcolumn}
\usepackage{bm}

\usepackage{color}

\usepackage{mathptmx}
\usepackage{latexsym}
\usepackage{keyval}
\usepackage{fancybox}

\newcommand{\be}{\begin{equation}}
\newcommand{\ee}{\end{equation}}
\newcommand{\bee}{\begin{eqnarray}}
\newcommand{\eee}{\end{eqnarray}}
\newcommand{\hs}{\hspace{0.8cm}}

\newcommand{\footnoteremember}[2]{\footnote{#2}\newcounter{#1}\setcounter{#1}{\value{footnote}}}

\newcommand{\footnoterecall}[1]{\footnotemark[\value{#1}]} 

\newcommand{\V}{\overrightarrow}
\newcommand{\M}{\bm}

\newcommand{\Vm}[1]{\V{#1}}
\newcommand{\VmDOT}[1]{\dot{\Vm{#1}}}
\newcommand{\VOm}[1]{\V {O #1}}

\newcommand{\transp}[1]{{#1}^T}

\newcommand{\Fij}{\V F_{ij}} 	
\newcommand{\FELA}{\V F^{ela}} 	
\newcommand{\FVIS}{\V F^{vis}}	
\newcommand{\FREM}{\V F^{rem}}	
 \newcommand{\FEXT}{\V F^{ext}}	
 
\newcommand{\FELADot}{ \dot {\V F^{ela}_{ij}} }
\newcommand{\FEXTDot}{\dot {\V F^{ext}_{i}} }

\newcommand{\Vd}{\V \delta} 	
\newcommand{\VdDot}{\dot  {\V {\delta}}}
\newcommand{\Vn}{{\V n}}
\newcommand{\VnDot}{{\dot {\V n}}}

\newcommand{\dn}{\delta^n}

\newcommand{\aDot}{\dot a}
\newcommand{\ad}{a_{\dn} }
\newcommand{\ah}{a_{h} }

\newcommand{\Ma}{\M \alpha}
\newcommand{\MaDot}{\M {\dot \alpha} }
\newcommand{\Mi}{\M 1}
\newcommand{\ME}{\M E}
\newcommand{\MEDot}{\dot {\ME}}
\newcommand{\MZ}{\M Z}

\newcommand{\MG}{\M G}

\begin{document}

\preprint{}

\title{\textit{Soft Dynamics} simulation:\\ 
2. Elastic spheres undergoing a T1 process in a viscous fluid}

\author{Pierre Rognon}
\author{Cyprien Gay}

\email{cyprien.gay@univ-paris-diderot.fr}

\affiliation{%
Centre de Recherche Paul Pascal, CNRS UPR~8641 - Av. Dr. Schweitzer, Pessac, France\\
Mati\`{e}re et Syst\`{e}mes Complexes, Universit\'{e} Paris-Diderot - Paris 7, CNRS UMR~7057 - Paris, France
}%

\date{\today}

\begin{abstract}
Robust empirical constitutive laws for granular materials 
in air or in a viscous fluid 
have been expressed in terms of timescales 
based on the dynamics of a single particle.
However, some behaviours such as viscosity bifurcation or shear localization, 
observed also in foams, emulsions, and block copolymer cubic phases,
seem to involve other micro-timescales which may be related 
to the dynamics of local particle reorganizations.
In the present work, we consider a T1 process as an example of a rearrangement.
Using the \textit{Soft dynamics} simulation method 
introduced in the first paper of this series,
we describe theoretically and numerically 
the motion of four elastic spheres in a viscous fluid.
Hydrodynamic interactions are described at the level of lubrication
(Poiseuille squeezing and Couette shear flow)
and the elastic deflection of the particle surface
is modeled as Hertzian.
The duration of the simulated $T_1$ process
can vary substantially as a consequence of minute changes
in the initial separations, consistently with predictions.
For the first time, a collective behaviour is thus found
to depend on another parameter than the typical volume fraction in particles.
\end{abstract}

\maketitle

\section{Introduction}\label{Sec:Intro}

Many materials are made of particles in a surrounding fluid. Among them
foams, emulsions, granular matter, colloidal suspensions and micro
gels are of daily use. A great deal of research revealed their complex behaviors including
elastic, plastic and viscous characters~\cite{Weaire01,Coussot05,Stickel05,Wyss07}. 
This complexity results from the wide range of particle properties and particle
interactions involved. Great hints to comprehensive rheological models were obtained by
considering the dynamics of a single particle. Thus emerged the time $\sqrt{m/RP}$ for a
single grain of mass $m$,
accelerated by the normal stress $P$ (force $P\,R^2$),
to move over a distance comparable 
to its own size $R$~\cite{GDR04, Dacruz05a,Forterre08}, 
the time $\eta/P$ for a grain immersed in a fluid 
of viscosity $\eta$ subjected to the same normal stress~\cite{Cassar05}, 
and the relaxation time $\eta\,R^2/\sigma$ for a bubble or a droplet
with surface tension $\sigma$ 
in a viscous fluid~\cite{Durian95,Durian97,Sollich97}. 
The effective viscosity was expressed 
as an empirical function of these microscopic timescales~\cite{Cassar05}, 
thereby providing robust scaling expressions
for various properties of grains~\cite{Jop06,Rognon07PF,Rognon08JFM}
and bubbles~\cite{Tewari99,Gopal99, Gardiner00a,Gardiner99,Gardiner05,Kern04}.
%

Nevertheless, particulate materials exhibit 
some uncommon rheological properties 
which seem to involve other timescales.
Oscillatory shear experiments~\cite{Wyss07}, and more generally the
delayed adaptation of the shear rate to a sudden change in the applied
stress~\cite{Dacruz02,Coussot02c,Rouyer03,Coussot06,Eiser00a,Eiser00b,Pailha08} 
reveal long
internal relaxation processes. Other observations such as a critical shear rate below
which no homogeneous flow exists 
~\cite{Coussot02c, GDR04, Dacruz05a, Cassar05, Rognon08JR}, or the coexistence of
liquid and solid regions 
(shear localization, shear banding, cracks) 
in emulsions~\cite{Coussot02b,Becu06}, 
foams~\cite{Debregeas01,Kabla03,Janiaud06,Janiaud07}, 
wormlike micelles~\cite{Salmon03,Becu07} 
and granular materials~\cite{Dacruz05a,Huang05,Mills08,Rognon08JFM} also point to a
complex internal dynamics. Usually, this internal dynamics is qualitatively understood
as the competition between external solicitations that the particles experience and their
ability to move within their neighborhood~\cite{Coussot02c}. 
Such a mechanism is the core of the definition of the jamming transition in glassy
systems, which is a subject of intense
debate~\cite{Isa06,Vanhecke07,Lu08,Cates08,Olsson07, Heussinger09}.
Getting new insights into reorganization
micro-timescale should therefore clarify the origin of such properties 
and should also provide useful hints to refine and generalize 
existing models of the material response.

A common reorganization process is the separation of particles while other particles
approach and fill the void. 
When they involve four particles, these events, usually refered
to as $T_1$ processes when dealing with foams and emulsions, 
occur in deformed regions at a frequency 
proportional to the deformation rate (see
for example~\cite{Princen83,Okuzono93,Earnshaw94,Jiang99,Debregeas01,Cohen01,Kabla03,
Gopal03, Dennin04,Vincent06}). They relax some stress and dissipate some energy.
The relation between the duration of a $T_1$ and the local stress
is thus expected to affect the rheological behaviour of the material.
For dry foams, the $T_1$ dynamics has been shown to result 
from the surface tension and the surface viscosity~\cite{Durand06}. 
The stretching ability of the interfaces 
avoids the need to squeeze violently the fluid 
between the approaching bubbles,
and its viscous dissipation is thus negligible. 
By contrast, the $T_1$ dynamics is less well described in wet foams 
or other less concentrated systems. 
A comprehensive description of their dynamics requires
a careful description of the particle interaction. 
For instance, visco-elastic and even adhesive properties of particles
were shown to be important~\cite{Lacasse96a,Lacasse96b,Besson07}.

\begin{figure}[!b]
\begin{center}
\resizebox{1.\columnwidth}{!}{%
\includegraphics{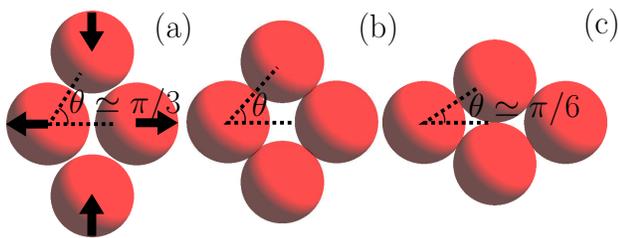}
}
\end{center}
\caption{Schematic representation of a $T_1$ process with four in-plane spheres.  Due to
the applied forces, the group of four particles swap neighbours. Two particles separate
while two other particles establish contact. Meanwhile, the other particle pairs
reorient, as shown by the evolution of angle $\theta$ 
from about $\frac{\pi}{3}$ to about $\frac{\pi}{6}$.
}
\label{Fig:T1}
\end{figure}

In this paper, we show that squeezing the liquid between close particles 
(here in three dimensions)
can give rise to long relaxation times. 
To this aim, we do not focus on a specific material which
would include the interaction between solid grains, bubbles, droplets
or colloidal particles. 
Rather, we address the ubiquitous situation of elastic-like particles
in
a Newtonian fluid. We consider a simple system of in-plane spheres undergoing a
$T_1$ process, as depicted on Fig.~\ref{Fig:T1}. We discuss under which circumstances
a $T_1$ process should indeed occur
and (if it {\em does} occur) the relative
contribution to the dynamic of the normal approach and separation~\textit{versus} the
tangential sliding of particles~\footnoteremember{footnote:no_rotation}{As the particle
configuration we consider 
is symmetric, it is not necessary to include particle rotation at this stage:
it will be introduced in a forthcoming paper.}. 
In three dimensions, for a dry foam or a concentrated emulsion, 
such a $T_1$ process with four topologically active particles 
can in fact be decomposed into two topologically simpler processes 
involving five particles. 
However, whatever the exact process, 
the dynamics will still involve normal motions and tangential sliding 
(as well as rotation in general). 
Because normal motions are stronger, as we show below,
we believe that no essential new phenomenon 
will emerge from other reorganization processes 
as compared to the time scale evidenced in the present work.

This paper is the second of a series which presents the physics of materials made of
close-packed elastic-like particles immersed in a viscous fluid. In the first
paper~\cite{Rognon08EPJE}, we focused on the normal separation of two particles, and we
showed that the flow between their close surfaces interplays with the particle
deformation in a non-trivial manner. As this feature was ignored so far in existing
discrete element simulations such as Molecular Dynamics~\cite{Cundall79} (for elastic
grains without a surrounding fluid) for Stokesian Dynamics~\cite{Durlofsky87}
(for non-deformable grains in a viscous fluid), we are introducing a new simulation
method, named \textit{Soft-Dynamics}, to account for it. 
In this paper, we include the
tangential interaction, and we provide the main steps of
the implementation of the \textit{Soft Dynamics} method 
for the present context. This will serve as an introduction to
the principle of larger scale simulations with this new method, which will include both
particle rotation and boundary conditions, and which should constitute a promising tool
for investigating the collective behaviors of many complex materials.

As we shall see, the geometry addressed in the present paper,
although rather symmetric (the centers of the three dimensional particles
are arranged within a plane, at the vertices of a losange),
proves sufficiently rich
to reveal how minute changes in the system configuration
have an essential influence upon its dynamics.

\section{Modelling particle interactions}\label{Sec:Interactions}

When addressing the question of a $T_1$ process
between elastic spheres in a viscous fluid, 
see Fig.~\ref{Fig:T1}, most of the interactions
have already been described in the first paper~\cite{Rognon08EPJE}.
The only new feature is particle sliding
and, correspondingly, tangential forces.
Hence, quantities such as viscous friction coefficients
or spring constants are now tensorial. 
We express these interactions in the present section.
Let us recall that we deal with three dimensional particles.

\subsection{Pairwise interactions}

As discussed in detail in the first paper~\cite{Rognon08EPJE},
because we consider rather dense systems
where each particle is close to several other particles
(surface-to-surface gap much smaller than the particle size),
we simply discard long-range, many-body interactions~\footnoteremember{footnote:long_range}{ As usualy done in the
Stokesian-Dynamic~\cite{Durlofsky87}, long range many-body hydrodynamic interactions can
be included in the Soft-Dynamic method. While not relevent for the system discussed here,
it must be included to simulate loose configurations as well as material with sparse
clusters. A detailed presentation of such interactions can be found in
Ref.~\cite{Abade07}. 
}.
Furthermore, under such thin gap conditions,
the interacting region between particles
is much smaller than the particle size
and the interactions between a particle and its neighbours 
are mostly independent from each other 
and can therefore be treated as a sum of pairwise interactions.

Particle $i$ is subjected to some force $\Fij$
by its neighbouring particles $j$,
which can be decomposted into
{\em (i)} the local pressure field in the fluid 
that results from a viscous lubrication interaction 
and {\em (ii)} a remote interaction 
(such as a damped electrostatic interaction,
steric repulsion, van der Waals interactions,
disjoining pressure, etc):
\be
\Fij = \FVIS_{ij}+\FREM_{ij}
\label{Eqn:balance_contact_ext} 
\ee

\subsection{Particle surface deflection}

A fraction of the above force $\Fij$ exerted by particle $j$
transits through a small portion of the surface of particle $i$
and deflects it elastically.

In practice, the viscous component $\FVIS$ of the force
is entirely transmitted by the surface of particle $i$.
The effect of the remote component $\FREM$ is more subtle.
Electrostatic forces between surface charges
act upon the surface and contribute entirely to the elastic deflection.
By contrast, Van der Waals interactions
also act directly within particle $i$.
However, most of such interactions occur
within a depth comparable with the inter-particle gap,
which is always much smaller than the depth of the region
that is deformed elastically
(see paragraph \ref{subsection:elastic_force} below).

Hence, for simplicity, 
it is reasonable to assume that the total force $\Fij$
between both particles entirely contributes
to the elastic surface deflection:
\be
\FELA_{ij} \simeq \Fij
\label{Eqn:balance_contact_int} 
\ee
The expression of $\FELA_{ij}$
in terms of the corresponding surface deflection
is discussed in paragraph \ref{subsection:elastic_force} below.

\subsection{Force balance for each particle}

The sum of all forces applied to particle $i$,
both the external force $\FEXT_i$
and the pairwise forces $\Fij$
is equal to the mass $m_i$ times the acceleration.
This is assumed to vanish
due to the dominant effect of the fluid viscosity over inertia:
\be
\FEXT_{i} + \sum_j \Fij = m_i\,\ddot{{\V X}_i}=\V 0
\label{Eqn:Part_equ}
\ee
where $\FEXT_{i}$ is an external force
acting on grain $i$ (such as gravity)
and where the sum runs over the neighbours of particle $i$.
In principle, there is another equation,
similar to Eq.~(\ref{Eqn:Part_equ}),
for the torques applied to particle $i$.
But as mentioned earlier~\footnoterecall{footnote:no_rotation},
this is not needed for the present symmetric $T_1$ configuration
such as that of Fig.~\ref{Fig:T1}.

The {\em Soft Dynamics} method~\cite{Rognon08EPJE}
simulates the dynamics of such a system, 
determined by the system of 
Eqs.~(\ref{Eqn:balance_contact}) for all interactions and 
Eqs~(\ref{Eqn:Part_equ}) for all particles $i$. 
In the present work, for simplicity, 
we omit the remote interactions
in Eq.~(\ref{Eqn:balance_contact})
as we did before~\cite{Rognon08EPJE}.

In order to specify the elastic and viscous forces,
let us now describe the geometry and the kinematics
of the interacting region between a pair of neighbouring particles.

\subsection{Contact geometry and kinematics} \label{Sec:Kinematics}

\begin{figure}[!htb]
\begin{center}
\resizebox{1.\columnwidth}{!}{%
\includegraphics{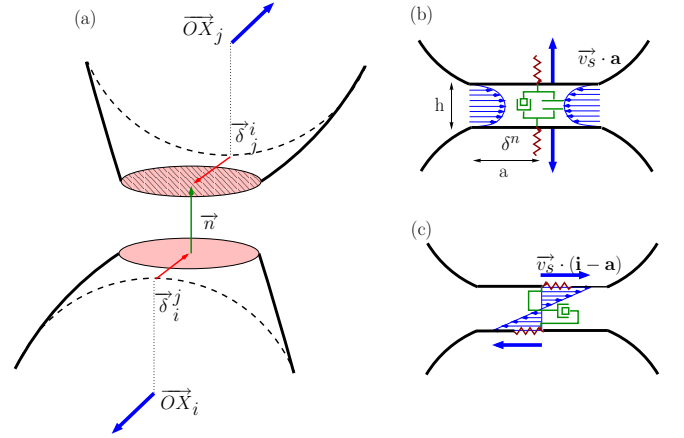}
}
\end{center}
\caption{ Model of interaction for two elastic spheres in a viscous fluid. (a) elastic
deflection of the surfaces (in traction); (b) normal dissipation due to Poiseuille flow in
the gap; (c) tangential dissipation due to the Couette flow.
The force is transmitted from a particle to another through the fluid and through a
possible remote force. Such a system behaves like a Maxwell fluid (a dashpot and a spring
in series). The effective friction is a function of the gap $h$ and of the size $a$ of the
surface through which the force is transmitted.
}
\label{Fig:contact}
\end{figure}

Let $i$ and $j$ denote two interacting particles,
as depicted on Fig.~\ref{Fig:contact}.
As compared to the first paper,
the positions of the particles centers are now vectors, 
labeled $\VOm{X_i}$ and $\VOm{X_j}$,
and $\Vm{X}_{ij}=\VOm{X_j}-\VOm{X_i}$ is the center-to-center vector.
The deflections of the particle surfaces are also vectors, 
labeled $\Vd_i^j$ and $\Vd_j^i$.
Since all particles are identical
and since the (lubrication) forces are pairwise and act locally,
facing deflections are symmetric: $\Vd_i^j+\Vd_j^i=0$.
Thus, for simplicity, we shall use the total deflection 
$\Vd_{ij} = \Vd_i^j -\Vd_j^i$
for each pair of interacting particles.
The unit vector normal to the contact can be expressed as 
\be
\label{Eqn:Vn}
\Vn_{ij} = \frac{\Vm{X_{ij}}-\Vd_{ij}} {\vert\Vm{X_{ij}}-\Vd_{ij}\vert}
\ee
The gap $h_{ij}$ between both particle surfaces 
depends on both the center-to-center vector $\Vm{X_{ij}}$
and the total deflection $\Vd_{ij}$:
\be
\label{Eqn:h}
h_{ij} =  \vert \Vm{X_{ij}}-\Vd_{ij} \vert  -2R , 
\ee
Similarly, the relative velocity of the material points
that constitute each particle surface, $\Vm{v}_s$, 
involves the translation velocity of the particles
(as already mentioned~\footnoterecall{footnote:no_rotation}, 
the particles do not rotate in the present situation)
and the evolution of the surface deflection:
\be
\Vm{v}_s = \VmDOT{X}_{ij} -\VdDot_{ij}. \label{Eqn:vs} 
\ee

In order to specify viscous and elastic interactions,
we will need to deal with projectors and tensors.
We will use the symbol ``$\cdot$'' for the tensor product 
(contraction of one coordinate index), 
and $\transp{\Vm{u}}$ will denote the transposed of vector $\Vm{u}$. 
Hence, $\transp{\Vm{u}} \cdot\Vm{v} = \transp{\Vm{v}} \cdot\Vm{u}$ 
will be the scalar product of $\Vm{u}$ and $\Vm{v}$, 
and $\Vm{u} \cdot \transp{\Vm{v}}$ their outer product,
which is a tensor.
In particular, we will make use of tensor $\Ma$ 
defined as the projector onto the normal direction: 
\be
\label{Eq:projector_alpha}
\Ma = \Vn \cdot \transp{\Vn}
\ee

\subsection{Viscous force}

For a pair of close spheres,
as discussed earlier~\cite{Rognon08EPJE},
the fluid region that mediates most of the force
between both particles has a large aspect ratio, 
and the flow is essentially parallel to the solid surfaces:
the \textit{lubrication approximation} can be used
(see for example~\cite{Batchelor67}). 
As before, the fluid inertia is negligible (low Reynolds numbers)
and the viscous force $\FVIS$ acting on the surfaces 
depends linearly on their relative velocity $\Vm{v}_s$:
\bee 
\label{Eqn:viscous_force}
\FVIS &=& \MZ \cdot \Vm{v}_s \\
\label{Eqn:viscous_force_Z}
\MZ &=& \zeta \Ma + \lambda (\Mi - \Ma) \\
\label{Eqn:viscous_force_zeta}
\zeta &=& \frac{3\pi\eta a^4} {2h^3} \\
\label{Eqn:viscous_force_lambda}
\lambda &=& \frac{\pi\eta a^2}{h}
\eee
where the interparticle friction tensor $\MZ$
has two components (normal and in-plane),
expressed in terms of the unity tensor $\Mi$
and the projector $\Ma$ defined by Eq.~(\ref{Eq:projector_alpha}).
The normal viscous friction $\zeta$
is related to the Poiseuille flow induced 
by squeezing or pulling~\cite{Rognon08EPJE}
(see Fig.~\ref{Fig:contact}b),
while the in-plane friction coefficient $\lambda$
reflects the tangential motion (sliding) between both particles,
which generates a Couette (shear) flow in the gap
(see Fig.~\ref{Fig:contact}c).

\subsection{Elastic force}
\label{subsection:elastic_force}

Let us assume that the size $a$ (discussed in the next paragraph)
of the interacting region between particles $i$ and $j$ is known.
Then, as before~\cite{Rognon08EPJE},
the force depends linearly on the surface deflection,
but this time the relation is tensorial:
\bee 
\FELA &=& a \ME \cdot \Vd , \label{Eqn:elastic_force} \\
\ME &=& E \left( c^n \Ma + c^t \left(\Mi - \Ma \right) \right), \label{Eqn:E}
\eee
where the tensorial proportionality constant $\ME$
is essentially the (scalar) Young modulus $E$,
but incorporates geometrical constants on the order of unity
$c^n$ and $c^t$
for the normal and tangential responses, respectively.

The elastic response of bubbles and droplets were found to deviate from such a Hertz
elasticity~\cite{Lacasse96a,Lacasse96b}. Although they have no bulk elasticity, the
surface tension $\sigma$ confers them some elastic-like properties, and the elastic-like
force mainly depends on the deflection $\delta$, the size $a$ of the interacting region
and an effective Young modulus which scales like $\sigma/R$.

\subsection{Size of the interacting region}\label{Sec:size_a}

The size of the interacting region, again~\cite{Rognon08EPJE},
depends either on the gap thickness $h$ (Poiseuille regime)
when the particle surface is weakly deflected,
or on the normal force (Hertz regime)
when the particle surface can be considered planar.
In the first case, it can be expressed as $a \approx \sqrt{2\,R\,h}$.
In the second case, it is essentially
independent of the tangential force~\cite{Johnson85}
and can thus be expressed in terms of the normal deflection: 
$a \approx \sqrt{R \vert \delta^n \vert}
=\sqrt{R\,\vert\transp{\Vn } \cdot \Vd\vert}$.
As explained earlier~\cite{Rognon08EPJE},
for the purpose of the {\em Soft Dynamics} method,
we interpolate between both behaviours of $a$
in a simple manner:
\be
a(h,\delta^n) = \sqrt{R (2h+\vert \delta^n \vert)}.\label{Eqn:surface_contact}
\ee

\noindent The choice of this interpolation is not physically supported, but it does not
affect assymptotic behavior in both limits.

\section{Method of the soft-dynamics simulation}\label{Sec:System}

The Soft-Dynamics method aims at simulating the time evolution 
of a system of elastic particles and in a viscous fluid, 
such as depicted in previous sections. 
Like usual discrete simulation methods, 
the motion of each particle center 
results from the force balance, Eq.~(\ref{Eqn:Part_equ}). 
The specificity is that the interaction evolution 
results from the decomposition of the center-to-center distance
given by Eq.~(\ref{Eqn:h}).
As illustrated previously~\cite{Rognon08EPJE},
this generates a Maxwellian contact dynamics
through the combination of the elastic surface deflection
and the viscous response of the fluid in the gap: 
it is possible to move the center-to-center distance $\Vm{X}_{ij}$ while keeping constant the deflection $\Vd_{ij}$, and \textit{vice-versa}.
But as compared to a classical Maxwell behaviour,
the elastic element does always behave linearly
(Hertzian contact in the strong deflection regime),
and the viscous element does not have a constant value,
as it depend on the geometry of the gap,
see Eqs.~(\ref{Eqn:viscous_force}-\ref{Eqn:viscous_force_lambda}).

The Soft-Dynamics method consists in calculating the rate of change
of all center positions $\Vm{OX_i}$ 
and all gap deflections $\Vd_{ij}$ 
as a function of their current values, 
and integrating them over a small time step.

\subsection{Equations of motion} \label{Eqn:dynamics}

The system satisfies one equation per interaction,
namely Eq.~(\ref{Eqn:balance_contact}),
and one equation per particle, namely Eq.~(\ref{Eqn:Part_equ}).
We shall now see how it is possible to derive equations of motion.
For this, we need to express the unknowns
velocities $\VdDot_{ij}$ and $\VmDOT{X}_{ij}$
in terms of the current state of the system.

From Eqs.~(\ref{Eqn:Vn}), (\ref{Eqn:h})
and~(\ref{Eqn:surface_contact}),
it appears that the size $a$ of the interacting region
can be expressed as a function of $\Vm{X_{ij}}$ and $\Vd_{ij}$.
It then follows from Eqs.~(\ref{Eqn:Vn}) 
and~(\ref{Eqn:elastic_force})
that the elastic force $\FELA$
can also be expressed as a function of $\Vm{X_{ij}}$ and $\Vd_{ij}$:
\be
\FELA=\FELA(\Vm{X_{ij}},\,\Vd_{ij})
\ee
As a result, its time-derivative $\FELADot$
can be expressed as a sum two terms:
one of them is linear in $\VmDOT{X}_{ij}$
while the other is linear in $\VdDot_{ij}$.
The (tensorial) coefficient of each of these two terms
is a function of the current system configuration,
{\em i.e.}, of all particle and gap variables 
$\VOm{X_i}$ and $\Vd_{ij}$.
Now, it follows from Eqs.~(\ref{Eqn:vs}),
(\ref{Eqn:viscous_force}) and~(\ref{Eqn:balance_contact})
that $\VdDot_{ij}$ is an affine function of $\VmDOT{X}_{ij}$:
\be
\VdDot_{ij}=\VmDOT{X}_{ij}
+ \MZ_{ij}^{-1} \cdot \left( \FELA_{ij} - \FREM_{ij} \right) \label{Eqn:delta_dot}
\ee
where $\MZ_{ij}$, $\FELA_{ij}$ and $\FREM_{ij}$
depend on the current system configuration.
Hence, $\FELADot$ can be expressed as an affine function of $\VmDOT{X}_{ij}$:
\be \label{Eqn:FELADot_affine}
\FELADot = \MG_{ij} \cdot  \VmDOT{X}_{ij} - \Vm{b}_{ij}
\ee
where the coefficients $\MG_{ij}$ and $\Vm{b}_{ij}$
depend only on the current system configuration.
The detailed calculation of these coefficients
is provided in Appendix~\ref{App:Xdynamics}.

From this, the time derivative of Eq.~(\ref{Eqn:Part_equ})
yields a system of equations for the particle center velocities.
The equation that corresponds to particle $i$ reads:

%
\be \label{Eqn:system}
\sum_j \left\lbrace  \MG_{ij} 
\cdot (\VmDOT{OX}_{j}-\VmDOT{OX}_{i})  \right\rbrace 
 =\sum_j  \Vm{b}_{ij} - \FEXTDot.
\ee
where the sums run over all neighbours of particle $i$.

Note that because $\MG_{ji}=-\MG_{ij}$
and $\Vm{b}_{ji}=-\Vm{b}_{ij}$,
and if we assume that the sum of all external forces vanishes,
\be \label{Eqn:zero_external_force}
\sum_i \FEXT_i = 0,
\ee
then the sum of Eqs.~(\ref{Eqn:system}) for all particles $i$ vanishes.
In other words, these vector equations are not independent: 
one of them must be replaced, for instance, 
by the condition that the average particle velocity is zero:
\be \label{Eqn:referential}
\sum_i \VmDOT{OX}_{i} = 0
\ee

Let us consider the system of Eq.~(\ref{Eqn:referential})
(or a similar one)
together with Eqs.~(\ref{Eqn:system}),
taken for all particles $i$ except one.
This system of equations can be inverted
to obtain the particle center velocities $\VmDOT{OX}_{i}$.
The gap velocities $\FELADot$
are then calculated from Eq.~(\ref{Eqn:FELADot_affine}).

\subsection{Choice of a numerical step}

Gaining the center velocity $\VmDOT{OX_i}$ requires to solve the linear system (\ref{Eqn:system}). Standard and efficient procedures are available to inverse it. We used a second order Newtonian scheme for the numerical integration of particle position and as well as deflections. 
A typical time in the problem is the Stokes time $\tau$
taken by a single particle submitted to a typical force $F$ 
to move over a distance $R$ in a fluid with viscosity $\eta$,
see Eq.~(\ref{Eq:Stokes_time_tau}) below.
The numerical time step is set to $10^{-3}$ in units of $\tau$ 
for all simulations. 
Other numerical schemes, such as Runge Kutta method,
should make simulations faster. 
Furthermore, a study of the optimal required time step 
will be necessary when dealing 
with significantly more than only four particles.

\begin{figure}[!b]
\begin{center}
\resizebox{1.0\columnwidth}{!}{%
\includegraphics{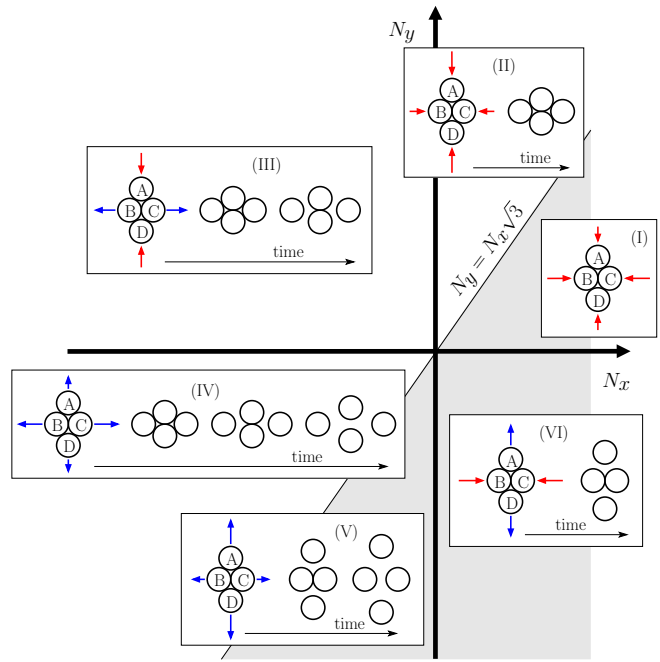}
}
\end{center}
\caption{Schematic evolution of four particles subjected to external forces. 
Force $N_x$ is horizontal and acts on particles $B$ and $D$. 
Force $N_y$ is vertical and acts upon particles $A$ and $C$. 
Both $N_x$ and $N_y$ can be either compressive ($>0$) or tensile ($<0$). 
Regimes (I) and (II) correspond to compressive forces. 
In regime (I), the configuration remains mostly unaltered. 
By contrast, a topological rearrangement ($T_1$ process) 
occurs when $ N_y \gtrsim \sqrt{3} N_x$, which corresponds to region (II). 
When $N_x$ or $N_y$ is tensile, the four beads do not remain together, 
as can be seen from the time evolutions sketched for regimes (III)-(VI). 
On the whole, a $T_1$ process always occurs when $N_y \gtrsim \sqrt{3} N_x$ 
(regimes II, III and IV, white region). 
It is followed by particle separation when $N_x$ 
is tensile (regimes III and IV).
By contrast, no $T_1$ process occurs when $N_y \lesssim \sqrt{3} N_x$ (regimes I, V and VI, light grey region).
}
\label{Fig1}
\end{figure}

\section{$T_1$ dynamics}

Let us now use the Soft-Dynamics method to simulate a single $T_1$ process. 
The system is depicted on Fig.~\ref{Fig1}: 
initially, particles $B$ and $C$ are aligned horizontally, with a small gap $h_0$, 
while particles $A$ and $D$ are aligned vertically.
The diagonal gaps (between $A$ and $B$, etc) have thickness $h_0$ too.

A horizontal force $N_x$ is applied on particles $B$ and $C$ 
while a vertical force $N_y$ is applied on $A$ and $D$. 
Various evolutions are possible depending on these two forces, 
which may or may not give rise to a $T_1$ process (see Fig.~\ref{Fig1}). 
Basically, a $T_1$ occurs only if the interaction between particles $B$ and $C$ is
tensile.
The criterion for the occurrence of a $T_1$ process will be derived below,
as well as a scaling for its dynamics.
The duration of a $T_1$ will then be measured from the simulation.

\subsection{Theoretical predictions} \label{Sec:theory}

The $T_1$ process, which consists in a separation 
of the horizontal pair of particles ($BC$)
and an approach of the vertical pair of particles ($AD$),
implies some sliding of the diagonal pairs (see Fig.~\ref{Fig:T1}).

At the early stages of the process, when $\theta\approx\pi/3$,
the external forces $N_x$ and $N_y$
can be expressed in terms of the normal forces
in the horizontal ($N_h$) and diagonal ($N_d$) pairs of particles,
and in terms of the sliding force $S_d$ in the diagonal pairs:
\bee
\label{Eq:Nx_Nh_Nd_Sd}
N_x&=&N_h+2\frac12\,N_d-2\frac{\sqrt{3}}{2}\,S_d \\
\label{Eq:Ny_Nd_Sd}
N_y&=&2\frac{\sqrt{3}}{2}\,N_d+2\frac12\,S_d
\eee

In fact, as we shall now see, the tangential force $S_d$
is much smaller than the normal forces.
To show this, let us first notice
that the tangential velocity is related
to the angle $\theta$ defined on Fig.~\ref{Fig:T1}:
$v_t\simeq -R\,\dot{\theta}$.
In the Poiseuille regime, the particles surfaces
are weakly deflected and the horizontal and diagonal gaps
are related to angle $\theta$ through
$R+\frac12\,h_h=(2R+h_d)\,\cos\theta$.
Hence, the gap variations obey
$\frac12 \dot{h}_h \approx \dot{h}_d\,\cos\theta-2R\,\dot{\theta}\,\sin\theta$,
{\em i.e.}:
\be
\frac12 \dot{h}_h \approx \dot{h}_d\,\cos\theta+2\,v_t\,\sin\theta
\ee
Let us now transform each term of the above equation 
by expressing it as a function of the corresponding 
normal or tangential force by using the appropriate friction coefficient
as defined by Eq.~(\ref{Eqn:viscous_force_Z}):
\be
-\frac12\,\frac{N_h}{\zeta} \approx -\frac{N_d}{\zeta}\,\frac12+2\,\frac{S_d}{\lambda}\,\frac{\sqrt{3}}{2}
\ee
The relative magnitude of friction coefficients $\zeta$ and $\lambda$ can be derived
from Eqs.~(\ref{Eqn:viscous_force_zeta})~and~(\ref{Eqn:viscous_force_lambda}):
\be 
\frac{\zeta}{\lambda} = \frac{3}{2}\left(\frac{a}{h} \right)^2.
\ee
\noindent where the size $a$ of the interaction region is given by Eq.~(\ref{Eqn:surface_contact}).
We thus have $\zeta/\lambda \approx R/h$ in the Poiseuille regime
and $\zeta/\lambda \approx R\delta^n / h^2$ in the Hertz regime.
Hence, except for very large gaps $h$ comparable to the particle size $R$,
the normal friction is much larger than the sliding friction: $\zeta\gg\lambda$.
It follows that 
\be
S_d\simeq\frac{1}{2\sqrt{3}}\,\frac{\lambda}{\zeta}\,(N_d-N_h)
\ee
can be neglected in Eqs.~(\ref{Eq:Nx_Nh_Nd_Sd}--\ref{Eq:Ny_Nd_Sd}).
Hence, the interaction force within the horizontal pair $BC$
depends only on the applied forces:
\be
\label{Eq:Nh_Nx_Ny}
N_h \approx N_x-\frac{1}{\sqrt{3}}\,N_y
\ee

This implies that, as pictured on Fig.~\ref{Fig1},
the gap will open and the $T_1$ will proceed
whenever $N_h$ is tensile, 
{\em i.e.}, when $N_y \gtrsim N_x\,\sqrt{3}$ (white region of the diagram).
By contrast, the particles will not swap neighbours
when $N_y \lesssim N_x\,\sqrt{3}$ (light grey region).

When $N_h$ is indeed tensile, we now wish to determine 
how long it takes for the horizontal pair of particles to separate.

The dynamics of such a normal motion 
was detailed in Ref.~\cite{Rognon08EPJE}.
Let us define the reduced force
\be
\label{Eq:kappa}
\kappa=\frac{|N_h|}{ER^2}
\ee
and the Stokes time 
\be
\label{Eq:Stokes_time_tau}
\tau=\frac{6\pi\eta R^2}{|N_h|}
\ee
With the force $N_h$ acting within the horizontal pair $BC$,
the initial configuration (gap $h_0$)
corresponds to the Poiseuille regime
if $h_0\gtrsim h_{HP}$
and to the Hertz regime
if $h_0\lesssim h_{HP}$, where
\be
h_{HP}=R\,\kappa^{2/3}
\ee
The corresponding rate of change of the gap~\cite{Rognon08EPJE}
can be expressed as:
\bee
\label{Eqn:hdot_poiseuille} 
\dot h &=& \frac{h}{\tau} \hs \text{(Poiseuille, $h>h_{HP}$)} \\
\dot h &=&  \frac{h^3}{\tau\,R^2} \kappa^{-\frac{4}{3}} 
\hs \text{(Hertz, $h<h_{HP}$)}. \label{Eqn:hdot_hertz}
\eee
Integrating these equations yields the typical time $\Delta$
required to achieve the separation of the horizontal pair $BC$ of particles
from an initial gap $h_0$ to a much larger gap $h_f \approx R$:
\bee
\label{Eqn:time_pois}
\Delta &\simeq& \tau\,\ln\left(\frac{h_f}{h_0}\right) \nonumber\\
&\approx& \tau \hs \text{(Poiseuille, $h_0>h_{HP}$)} \\
\Delta &\simeq& \tau\, \kappa^{\frac{4}{3}}\,
\left(\frac{R^2}{h_0^2} -\frac{R^2}{h_{HP}^2} \right)
+\tau\,\ln\left(\frac{h_f}{h_{HP}}\right)
\nonumber\\
&\approx& \tau\, \kappa^{\frac{4}{3}}\,\frac{R^2}{h_0^2}
\gg\tau
\hs \text{(Hertz, $h_0<h_{HP}$)} 
\label{Eqn:time_hertz}
\eee
Once the gap $h_h$ of the horizontal pair $BC$
becomes comparable to $R$,
the diagonal pairs such as $AB$ slide rather quickly 
(since their $\lambda\ll\zeta$),
and soon the gap $h_v$ of the vertical pair $AD$
becomes significantly smaller than $R$.
The time it then takes to reach the same value $h_0$
is again comparable to $\Delta$.

Hence, the order of magnitude 
given by Eqs.~(\ref{Eqn:time_pois}--\ref{Eqn:time_hertz})
for the time $\Delta$ 
is typically the expected order of magnitude 
for the duration of the entire $T_1$ process.
We will now test this prediction
by comparing it with the simulation results.

\subsection{Result from simulations}\label{Sec:Results}

We implement the {\em Soft Dynamics} method 
to simulate a $T_1$ process such as that depicted on Fig.~\ref{Fig1}, 
varying the two control parameters we pointed out above: 
the initial gap $10^{-3} < h_0/R < 0.8$ 
and the reduced force $10^{-4} < \kappa < 0.1$. 
For simplicity, there is no horizontal force  ($N_x=0$). 
The reduced force given by Eq.~(\ref{Eq:kappa})
is then equal to $ \kappa = |N_y|/ER^2\sqrt{3}$ 
and the Stokes time is $\tau = 6\pi\sqrt{3}\eta R^2/|N_y|$.

\begin{figure*}[!bht]
\begin{center}
\resizebox{1.8 \columnwidth}{!}{%
\includegraphics{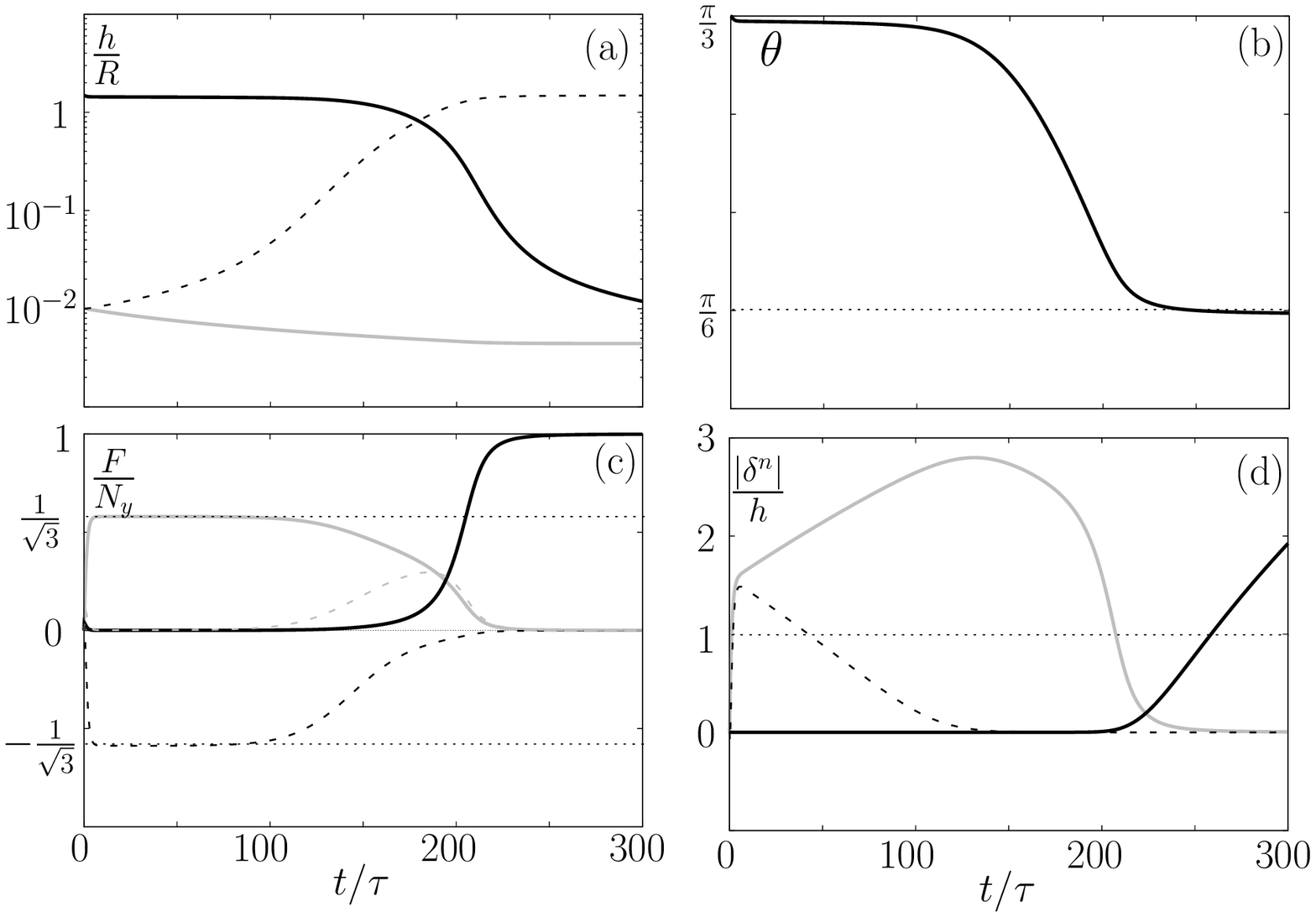}
}
\end{center}
\caption{ \label{Fig3}
Time evolution of various quantities in the course of a $T_1$ process, 
with parameters $\kappa = 3 10^{-3}$, $h_0= 10^{-2} R$ and $N_x=0$. 
The gap $h$ {\em (a)}, the normal and tangential forces {\em (c)} 
and the ratio $|\delta^n|/h$ {\em (d)}
are plotted for the vertical (solid black lines, $AD$), 
the horizontal (dotted black lines, $BC$) 
and the diagonal (solid grey lines, $AB$ etc) pairs of particles. 
On graph {\em (c)}, the dotted grey line 
represents the tangential force $S_d$
of a diagonal pair such as $AB$ 
(which is zero for the vertical pair $AD$ and horizontal pair $BC$). 
Graph {\em (b)} shows the angle $\theta$ 
such as defined on Fig.~\ref{Fig:T1}. 
On graph {\em (d)}, a pair of particles 
for which $|\delta^n|/h>1$ is in the Hertz regime.
If $|\delta^n|/h<1$, it is in the Poiseuille regime.
}
\end{figure*}

Figure~\ref{Fig3} displays the variations of several quantities
in the course of a $T_1$ process
with a given set of parameters 
($h_0 = 10^{-2}R$, $\kappa = 3.10^{-3}$). 
In order to avoid discontinuities in the simulation,
the force $N_y$ is increased from zero to its nominal value
within a time $\tau$, and remains constant thereafter. 
From a macroscopic point of view, 
for instance through the variation of the angle $\theta$, 
the system seems to be almost blocked ($\theta \approx \pi/3$) 
for a significant amount of time ($t\lesssim 100 \tau$). 
It then starts moving to reach its final configuration 
($\theta \approx \pi/6$)
where it remains thereafter ($t\gtrsim 250 \tau$).
During the ``blocked'' phase, the applied force $N_y$ 
is transmitted through the diagonal interaction such as $AB$, 
thereby inducing a tensile force $N_h\approx -N_y/\sqrt{3}$ 
in the horizontal pair $BC$. 
Hence, despite the overall ``blocked'' appearance of the system, 
the horizontal gap $h_h$ between particles $B$ and $C$
slowly increases from its initial value $h_0$. 
Correspondingly, the horizontal friction decreases.

The fast moving period starts as soon as this friction is low enough. 
Particles $B$ and $C$ then separate quickly 
while particles $A$ and $D$ in the vertical pair approach each other,
thereby giving rise to sliding friction on the diagonal interactions. 
As particles $A$ and $D$ approach, 
the corresponding gap $h_v$ decreases and the friction increases. 
This approach then slows down. 
Thus, although the system keeps moving, 
it {\em appears} to reach a new ``blocked'' configuration, 
with no more sliding or horizontal traction, 
but only a vertical compression.

\begin{figure*}[!htb]
\begin{center}
\resizebox{2.\columnwidth}{!}{%
\includegraphics{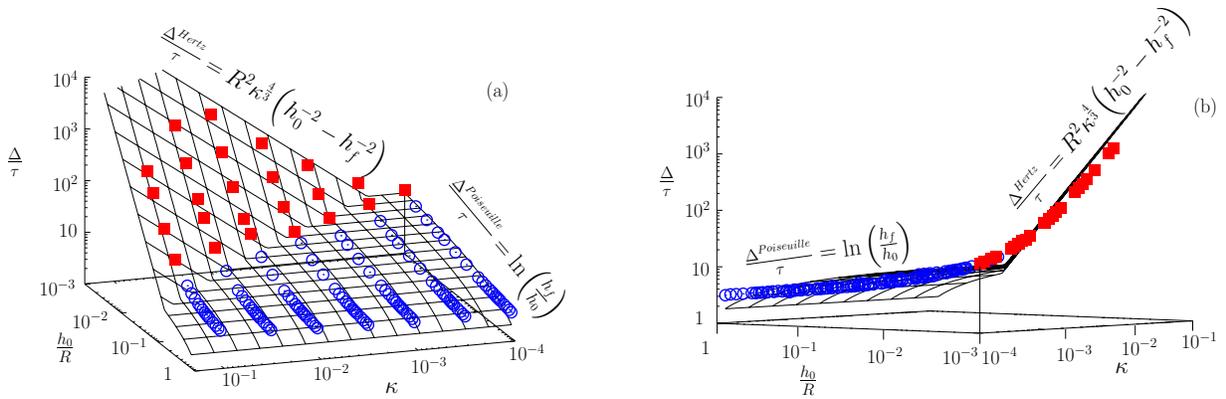}
}
\end{center}
\caption{
(Color online) Typical duration $\Delta$ of a $T_1$ process 
as a function of the initial gap $h_0$ 
and of the dimensionless applied force $\kappa$.
The data points were obtained 
through the Soft-Dynamics simulation presented here.
Blue open circles correspond to $T_1$s
where the horizontal pair has remained in the Poiseuille regime 
during the entire process.
Full red squares correspond to $T_1$s,
such as that represented on Fig.~\ref{Fig3},
whose horizontal (separating) pair
has been in the Hertz regime
for part of the time.
The surface is that defined 
by the theoretical model for both regimes 
(Eqs.~\ref{Eqn:time_pois}--\ref{Eqn:time_hertz} with $h_f=2.5R$).
}
\label{Fig4}
\end{figure*}

As the system subjected to a constant force keeps moving, 
we need to arbitrarily define the end of the $T_1$ process. 
Among various possible choices, 
we shall here consider that the $T_1$ process is completed 
when the vertical interaction transmits 
most of the applied force ($N_v = 0.99\,N_y$). 
The resulting duration of the $T_1$ process
is plotted on Fig. \ref{Fig4} 
(other criteria would yield similar results). 
The first observation is that, 
for the range of initial gaps and particle stiffnesses we consider, 
the duration of the $T_1$ 
is distributed over a wide range of time scales,
roughly between $3\tau$ and $10^3\tau$. 
Next, we observe that these results 
match our theoretical predictions reasonably:
\begin{itemize}
\item if the horizontal pair $BC$ is in the Poiseuille regime,
the $T_1$ duration $\Delta$ scales like $\tau\,\ln(R/h_0)$. 
It thus depends on particle radius, on the applied force 
and on the fluid viscosity through $\tau$,
as can be seen from Eq.~(\ref{Eq:Stokes_time_tau}),
and slightly on the initial gap through the logarithmic factor.
The $T_1$ duration is then just a few times larger than the Stokes time $\tau$;
\item if pair $BC$ is in the Hertz regime,
the $T_1$ duration $\Delta$ scales essentially like 
$\tau\,\kappa^{\frac{4}{3}}\left(\frac{R}{h_0}\right)^2$, 
which implies a much stronger dependence on $h_0$, 
and longer durations since the particles are soft. 
In this case, $\Delta$ can be much longer 
than the Stokes time $\tau$.
\end{itemize}
Note that in the latter case,
the separating pair of particles
leave the Hertz regime and enter the Poiseuille regime
in the late stages of separation ($h>h_{HP}$).
However, because the evolution is much slower in the Hertz regime,
see Eqs.~(\ref{Eqn:hdot_poiseuille}--\ref{Eqn:hdot_hertz}), 
these late Poiseuille stages 
contribute very weakly to the overall $T_1$ duration $\Delta$.

In summary, the numerical result for the duration of a $T_1$ process
presented on Fig.~\ref{Fig4} are compatible 
with Eqs.~(\ref{Eqn:time_pois}--\ref{Eqn:time_hertz}).
They demonstrate that the duration of a $T_1$
is hardly larger than the Stokes time $\tau$
given by Eq.~(\ref{Eq:Stokes_time_tau})
as long as the surface deflection
is small compared to the inter-particle gap (Poiseuille regime)
and thus depend mainly on the applied force,
on the fluid viscosity and on the particle size.
Remarkably, in the opposite regime 
where the deflection is larger than the gap (Hertz regime), 
the $T_1$ duration depends strongly
on the interparticle gap and can reach very large values,
as illustrated by Fig.~\ref{Fig4}.

\section{Conclusion: beyond volume fraction}\label{Sec:Ccl}

In this paper, we studied one of the simplest
reorganization processes for immersed, closed-packed, 
elastically deformable particles in a simple geometry.
We showed that the time needed for this process
results principally from the viscous flow of the fluid
into or out of the gap between pairs 
of almost contacting particles:
it is always mostly driven 
by the normal approach or separation,
while the role of tangential sliding is negligible.

We also showed that the time needed
can be {\em very long} when particles are close or soft
(more explicitely, when the gap is much thinner
than the particle surface deflection).
This is the central result of the present study
and, as we show below, it pleads towards
going beyond the sole usual volume fraction
to describe the state of a particulate material.

\subsection{Volume fraction and interparticle gap}

Let us consider four particles in a compact configuration
such as that on Fig.~\ref{Fig:T1}a
(angle $\theta\approx\pi/3$).
More precisely, let us consider two variants
of this configuration, with two different values
of the interparticle gap $h_0$,
say $h_0\simeq 10^{-2}\,R$ and $h_0\simeq 10^{-3}\,R$.
Let us now apply weak forces (say $\kappa\simeq 10^{-2}$).
In both situations,
because the force is weak and the gaps are small,
the center-to-center distances are almost identical.
Hence, both situations cannot be distinguished at first sight.

Yet it can be seen from Fig.~\ref{Fig4}
that the duration of the $T_1$ process
will then differ substantially.

Similarly, with a large, disordered assembly of grains,
it is anticipated that there will exist different situations
where the volume fraction is almost identical
but where a change in the typical value of the interparticle gaps
causes a dramatic alteration of the delay $\Delta$
after application of the stress
for the system to set into appreciable motion.

This conjecture will be tested in a future work,
using simulations with a large number of particles.

\subsection{Dilatancy and permeation}

In even larger samples of granular materials in a compact state,
it is anticipated that the need for some additional fluid
to enable reorganization processes (a phenomenon called dilatancy
and illustrated on Fig.~\ref{Fig:T1})
will become the main source of delay:
fluid from loose or particle-free regions 
needs to permeate through the granular material 
which behaves as a porous medium~\cite{Pailha08}.
Describing such a phenomenon requires
introducing the liquid pressure,
and is not included in our simulation so far.

\subsection{Towards other materials}

As such, the present work applies to soft, plain, elastic particles
(such as elastomer beads or latex particles)
immersed in a very viscous fluid.
We showed that a possible physical origin
for a delay in the system response
is the viscous flow in the thin gap 
between neighbouring particle surfaces.

In other materials, however,
other ingredients may also influence this delay
or even become dominant.
For instance, for objects enclosed in fluid interfaces
(vesicles, onions, bubbles, droplets, etc),
phenomena such as Marangoni effects,
surface viscosity,
the dynamics of surfactant adsorption
and Gibb's elasticity should play a role.
Finer phenomena should also be considered,
such as the hydrodynamics involved
either near a moving ``contact'' line between two such objects
or within Plateau borders.
By contrast, for solid grains, very different phenomena
may come into play, including solid friction.

For each of these phenomena,
a simplified yet realistic pairwise interaction law
will need to be expressed
and can then be included in the simulation rather easily.

\subsection{Perspectives}

The present study suggests that further investigations 
using the \textit{Soft-Dynamics} method with larger systems 
(including particle rotation as well as boundary conditions)
should provide interesting results,
not only with the present system of plain, 
elastic beads in a viscous fluid,
but also with different types of particle interactions.
By testing ideas such as the influence of the typical interparticle gap
(or other quantities if the interactions are different),
it should also provide hints for analytical modelling
beyond the role of the particle volume fraction.

\section*{Acknowledgments}

We gratefully acknowledge fruitful discussions with Fran\c{c}ois Molino
and with participants of the GDR 2983 Mousses (CNRS).
This work was supported 
by the Centre National de la Recherche Scientifique (CNRS),
by the Universit\'e Paris-7 Paris Diderot, 
and by the Agence Nationale de la Recherche (ANR-05-BLAN-0105-01).

\appendix

\section{Dynamics of particles}\label{App:Xdynamics}

In this Appendix, we deduce the particle dynamics, 
given by Eqs.~(\ref{Eqn:system}), 
from the physical model of interactions 
and the mechanical equilibria described in Sec.~\ref{Sec:System}. 
We start from the time derivative 
of the particle force balance given by Eq.~(\ref{Eqn:Part_equ}):

\be \label{Eqn:part_equ_dot}
\sum_j \FELADot + \FEXTDot = 0,
\ee

\noindent Let us express the above as a sum of 
{\em (i)} $\FEXTDot$ which is supposed to be known,
{\em (ii)} terms that are linear in the particle center velocities, 
and {\em (iii)} another term that is explicitly known 
from the current state of the system, 
{\em i.e.}, from $\Vm{X}_{ij}$ and $\Vd$. 
To this aim, using Eq.~(\ref{Eqn:elastic_force}),
let us express $\FELADot$ 
in terms of the partial derivative of $\FELA(a,\ME,\Vd)$:
\be \label{Eqn:Fela_Dot}
 \FELADot =\frac{a}{2}  \ME \cdot \VdDot  + \frac{a}{2} \MEDot \cdot \Vd + \aDot \frac{
\ME \cdot  \Vd } {2}   .
\ee

\subsection{Combination}

We can now easily express each terms of Eq.~(\ref{Eqn:Fela_Dot}) 
as a function of $\VmDOT{X}_{ij}$. 
The first term is directly given by (\ref{Eqn:vs}):
\bee \label{Eqn:TERM1}
  &&\frac{a}{2} \ME \cdot \VdDot = \M {G_1} \cdot  \VmDOT{X}_{ij} - \Vm{b}_1\\ 
&&\M {G_1} =\frac{a}{2} \ME\\
&&\Vm{b}_1 = \M {G_1}\cdot \Vm{v}_s
\eee
\noindent The second term involves the time derivative 
$\MEDot = E(c_n-c_t) \MaDot$
of the contact stiffness expressed by Eq.~(\ref{Eqn:E}). 
Using (\ref{Eqn:alphaDot}), it can be expressed as:
\bee\label{Eqn:TERM2}
&&a \frac{\MEDot}{2} \cdot \Vd = - \Vm{b}_2\\
&&\Vm{b}_2 =  -\frac{a E(c_n-c_t)}{2} \MaDot \cdot \Vd 
\eee

\noindent Note that this term vanishes for $c_n=c_t$.

The third term involves the time derivative of $a$, 
which we express through its partial derivatives: 
$ \aDot(\dn,h) = \ad \dot \dn  +  \ah \dot {h} $ 
(we used the notation $\ad = \frac{\partial a}{\partial \dn} $ 
and $\ah = \frac{\partial a}{\partial h}$ ). 
Replacing $\dot \dn$ and $\dot {h}$ 
by their respective expressions in terms of $\VmDOT{X}_{ij}$, 
Eqs.~(\ref{Eqn:dnDot}) and~(\ref{Eqn:hDot}) lead to:
\bee
&&\aDot \frac{\ME \cdot \Vd}{2} = \M{G_3} \cdot \VmDOT{X}_{ij} -
\Vm{b}_3\label{Eqn:TERM3}\\	
&&\MG_3  = \frac{\ad} {2} \ME \cdot \Vd \cdot  \transp{\Vn}\\
&&\Vm{b}_3 = \frac{\ME \cdot \Vd} {2} \left[\ad \left( \transp{\Vm{v}_s} \cdot \Vn  -
\transp{\Vd}\cdot \VnDot   \right)   - \ah \dot h \right]
\,\,\,\,\,\,\,\,\,\,\,\,
\eee

Finally, substituting the three results of Eqs.~(\ref{Eqn:TERM1}), (\ref{Eqn:TERM2})
and~(\ref{Eqn:TERM3}) into Eq.~(\ref{Eqn:Fela_Dot}) yields:
\be \label{Eqn:FINAL}
\MG \cdot \VmDOT{X}_{ij}  
= \V b  + \FELADot 
\ee

\noindent where $\MG=  \M {G_1}+ \M {G_3}$ and $\V b = \V b_1 + \V b_2 + \V b_3$ are two
explicit functions of $\Vd$ and $\Vm{X}_{ij}$ Then, for each particle $i$, summing on its
interacting particles $j$ and using the force balance (\ref{Eqn:Part_equ}) yields the
system of equation (\ref{Eqn:system}).

\subsection{Preliminary differentiations}

According to the definition of the normal vector, 
$\Vn = \frac{ \Vm{X_{ij}} - \Vd}{\vert \Vm{X_{ij}} - \Vd \vert}$, 
and to that of the associated projector, $\Ma=\Vn \cdot \transp{\Vn}$, 
we obtain their time derivatives:
\bee 
\VnDot &=& \left(\Mi-\Ma\right) \cdot \frac{ \Vm{v}_s }{ \vert \Vm{X_{ij}}  - \Vd \vert}
\label{Eqn:nDot}\\
\MaDot &=&   \VnDot \cdot \transp{\Vn} + \Vn \cdot \transp{\VnDot}\label{Eqn:alphaDot}
\eee
\noindent as two an explicit functions of $\Vd$ and $\Vm{X}_{ij}$. 
Indeed, according to Eq.~(\ref{Eqn:delta_dot}), 
$\Vm{v}_s$ can be expressed as a function of the elastic force: 
$\Vm{v}_s= \MZ^{-1}\cdot \left(\FELA-\FREM \right) $.

The evolution of the normal deflection  $\dn = \transp{\Vd} \cdot \Vn$ 
can be expressed as a function of $\VmDOT{X}_{ij}$ 
by using Eqs.~(\ref{Eqn:nDot}) and~(\ref{Eqn:vs}):
\bee
\dot {\overbrace{\dn}} &=& \transp{\VdDot} \cdot \Vn + \transp{\Vd} \cdot \VnDot \nonumber
\\
&=& \transp{\VmDOT{X}_{ij}} \cdot \Vn -\transp{\Vm{v}_s} \cdot \Vn +  \transp{\Vd} \cdot
\VnDot
\label{Eqn:dnDot}
\eee

Finally, from  Eq.~(\ref{Eqn:vs}), we deduce
the expression of the gap evolution $\dot h$ 
as a function of $\VmDOT{X}_{ij}$:
\be
\dot {h} = \left(\Vm{X_{ij}} - \Vd \right) \cdot \VnDot + \Vm{v}_s \cdot \Vn.
\label{Eqn:hDot}
\ee

\bibliography{T1}

\end{document}